\documentclass[twocolumn,granma]{svjour3} 

\usepackage{graphicx}
\usepackage{dcolumn}
\usepackage{bm}
\usepackage{color}
\usepackage{amsmath, amssymb, gensymb}
\usepackage{subfigure}
\usepackage{caption}
\usepackage[svgnames*,table,dvipsnames]{xcolor} 
\usepackage{ulem}

\begin{document}


\title{Packings of 3D stars: Stability and structure}

\author{Yuchen Zhao \and Kevin Liu \and Matthew Zheng \and Jonathan Bar\'es \and Karola Dierichs \and Achim Menges \and Robert P. Behringer}

\institute{Yuchen Zhao \and Jonathan Bar\'es \email{jb@jonathan-bares.eu} \and Robert P. Behringer \at Department of Physics \& Center for Nonlinear and Complex Systems, Duke University, Durham, North Carolina, USA \and Kevin Liu \at Julia R. Masterman Laboratory and Demonstration School, Philadelphia, Pennsylvania, USA \and Matthew Zheng \at North Carolina School of Science and Mathematics, Durham, North Carolina, USA  
\and Karola Dierichs \and Achim Menges \at Institute for Computational Design, University of Stuttgart, Stuttgart, Germany}

\date{Received: date / Revised version: date}
\maketitle


\begin{abstract}

We describe a series of experiments involving the creation of cylindrical packings of star-shaped particles, and an exploration of the stability of these packings. The stars cover a broad range of arm sizes and frictional properties. We carried out three different kinds of experiments, all of which involve columns that are prepared by raining star particles one-by-one into hollow cylinders. As an additional part of the protocol, we sometimes vibrated the column before removing the confining cylinder. We rate stability in terms of $r$, the ratio of the mass of particles that fall off a pile when it collapsed, to the total particle mass. The first experiment involved the intrinsic stability of the pile when the confining cylinder was removed. The second kind of experiment involved adding a uniform load to the top of the column, and then determining the collapse properties. A third experiment involved testing stability to tipping of the piles. We find a stability diagram relating the pile height, $h$, $vs.$ pile diameter, $\delta$, where the stable and unstable regimes are separated by a boundary that is roughly a power-law in $h~vs.~\delta$ with an exponent that is less than one. Increasing friction and vibration both tend to stabilize piles, while increasing particle size can destabilize the system under certain conditions.

\end{abstract}



\section{Introduction}
\label{intro}

Shape is one of the key features of a grain in determining mechanical
behavior of an aggregate
\cite{miskin2014_sm,athanassiadis2014_sm,smith2010_pre,Franklin2014_epl}. More
particularly, aggregates made from non-convex particles are an
emerging area of research that involves tuning material properties to
give specific functions to the macroscopic system
\cite{miskin2013_nm,SaintCyr2011_pre,brown2012_prl}. This is of high
interest not only in granular science
\cite{miskin2013_nm,Franklin2014_epl} but also for architectural
design \cite{dierichs2015_ad}. Non-convex particle geometries are one
main group within the overall area of designed aggregates, in addition
to double-non-convex hook-like, convex or even actuated particles
\cite{gravish2012_prl,Franklin2014_epl}. These approaches are
promising for future lightweight and reversible structural
applications. Indeed, such systems can be custom-designed for specific
mechanical properties \cite{miskin2014_sm,miskin2013_nm} and
architectural applications. Aggregate systems of particles can also be
programmable \cite{gershenfeld2012_fa}. The relevance of these
synthetic aggregate systems to construction is twofold: on the one
hand they can be fully re-configured and re-cycled; on the other hand
these structures can be functionally calibrated and graded according
to specific mechanical performance criteria.

Two of the most striking characteristics of non-convex designed
granulates, in terms of the granular system and possible design
applications are their ability to form vertical structures with a
$90^{\circ}$ angle of repose, and to sustain small tilting or loading
perturbations. The aim of this paper is to investigate, in a
laboratory setting, how geometry, material, and proportions, as well
as preparation and base properties affect the stability of vertical
columns. These data serve to inform the production of full-scale
architectural construction systems, which integrate mass-production
and on-site robotic construction \cite{dierichs2015_ad}.

Aggregate matter made of non-convex particles is also of interest in
terms of cohesive strength \cite{Franklin2014_epl}. This has been
investigated in studies of particles that are shaped like staples
\cite{gravish2012_prl,Franklin2014_epl,franklin2012_pt}. In this paper
we investigate the stability of star-shaped particle aggregates. Some
research has been done on the packing of such particles
\cite{miszta2011_nat,graaf2011_prl,jiao2009_pre} but little is known
about the stability of such aggregates. In the first part, we present
our experimental set-ups and the processes used to prepare cylindrical
packings. In the second part of this work, results on the intrinsic
stability of these aggregates is presented for different particle
dimensions and materials, for cylinder aspect ratios, and preparation
processes. The effect of vertical loading or tilting is also studied,
as well as destabilization due to the inclusion of beads in the
packing. Finally these results are discussed in the third part, and
rules for a better architectural design are extracted.



\section{Experimental set-up} \label{SecExpe}

\begin{figure}[htb!]
	\begin{center}
		\includegraphics[width=0.45\textwidth]{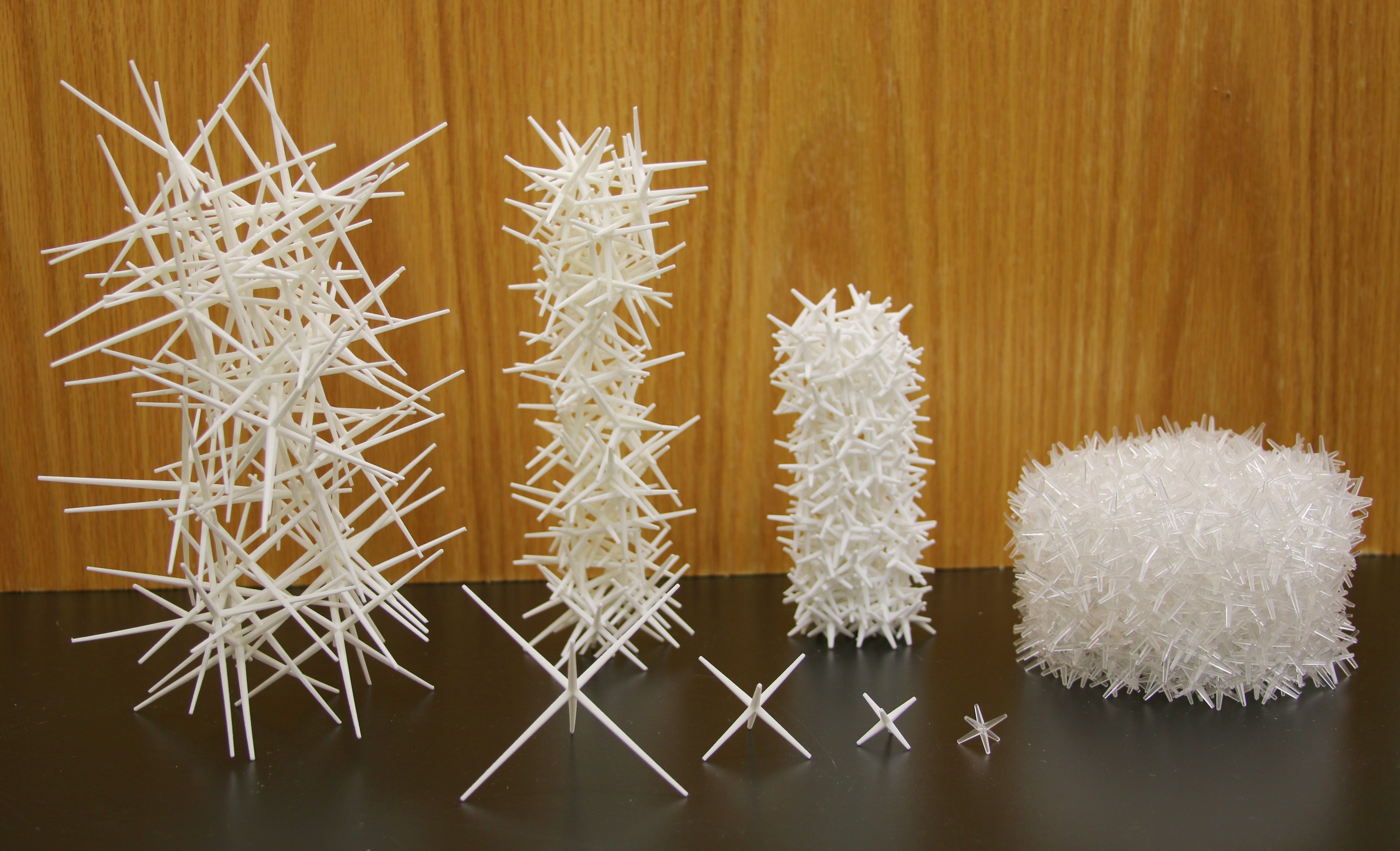}
	\end{center}
\caption{(color online) Non-convex particles used for
  experiments. From left to right: laser sintered white nylon PA2200
  particles with size $s$ $10$cm, $5$cm and $2.5$cm and cast acrylic
  particles of size $2$cm. Each star consists of six orthogonal square
  beams that taper from a thickness of $2$mm at the center of the
  particle to $1$mm at the tips.}
\label{figStar}
\end{figure}

We prepared cylindrical columns in several different ways and
subjected them to several different types of perturbations. We used
four kinds of stars, that were made from two different types of
materials, as shown in Fig.\ref{figStar}. Each star consisted of six
orthogonal beams with square cross section, that tapered from a
thickness of $2$mm at the center of the particle to $1$mm at the
tips. The end-to-end size varied from $2$cm to $10$cm. They are made
of either cast acrylic, which has a friction coefficient $0.4 \pm 0.1$
or of laser-sintered white nylon PA2200, which has a friction
coefficient $1.0 \pm 0.3$. In some cases, we also created
configurations that are a mixture of stars and beads. The latter were
made of acrylic and have a diameter of $9.5$mm and a mass of
$0.53$g. We summarize all particle properties in Tab.\ref{tabStar}.

\begin{table}[htb!]
	\begin{footnotesize}
		\begin{center}
			\begin{tabular}{|c|c|c|c|}
				\hline 
				maximum size & material & mass & friction coefficient \\ 
				(cm) &  & (g) &  \\
				\hline 
				$2$ & acrylic & $0.15$ & $0.4$ \\ 
				\hline 
				$2.5$ & nylon & $0.22$ & $1$ \\ 
				\hline 
				$5$ & nylon & $0.45$ & $1$ \\ 
				\hline 
				$10$ & nylon & $0.84$ & $1$ \\ 
				\hline 
			\end{tabular} 
		\end{center}
		\caption{Properties of the 3D non-convex particles.}
		\label{tabStar}
	\end{footnotesize}
\end{table}

\begin{figure*}[htb!]
	\begin{center}
		\includegraphics[width=0.9\textwidth]{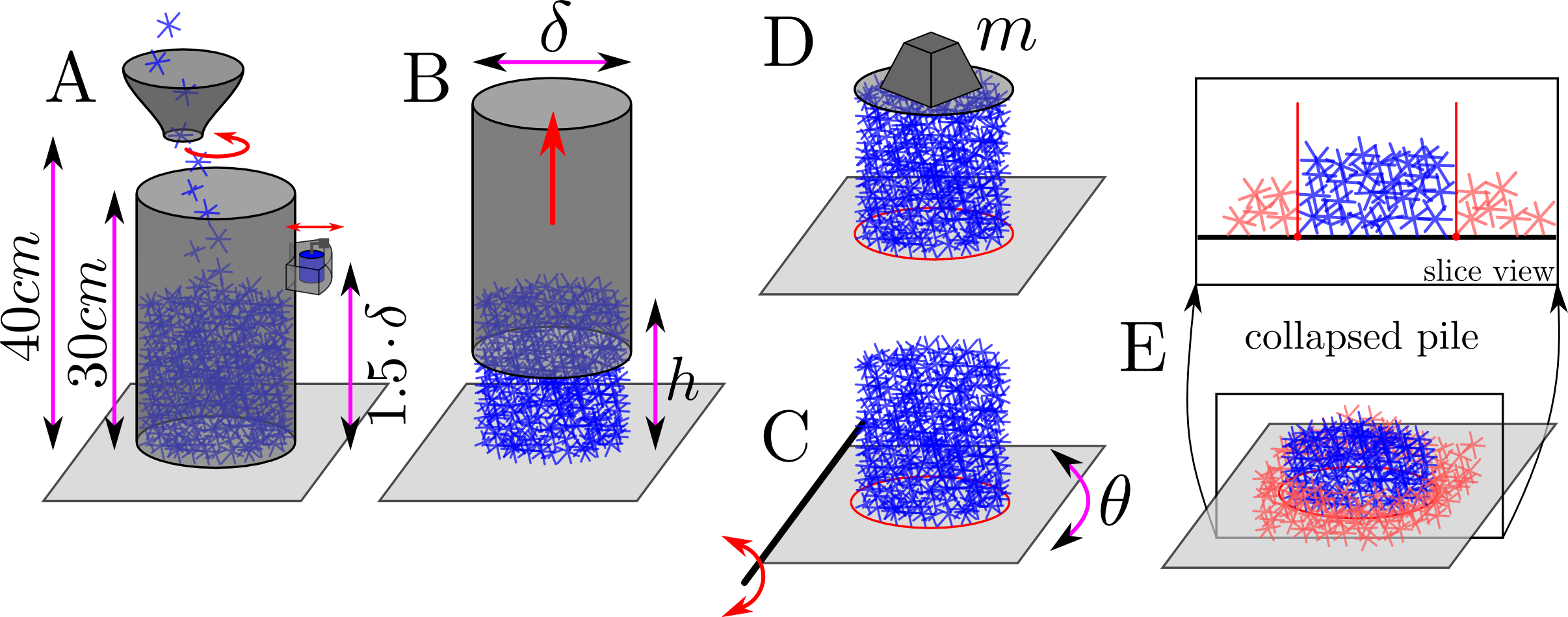}
	\end{center}
\caption{(color online) A: Non-convex particles are dropped one-by-one
  at a steady rate from an overhead hooper maintained at a constant
  height in a cylinder of diameter $\delta$ and height $30$cm. The
  system can be vibrated by an eccentric, driven by a DC motor that is
  attached to the cylinder at a height of $\sim 1.5 \cdot \delta$. B:
  Then the pile height, $h$ is measured and the cylinder is carefully
  and slowly removed by sliding lifting it vertically. C: If the pile
  is stable, this stability is tested by tilting the base of an angle
  $\theta$ until the system collapse. D: The stability is also tested
  by gradually increasing the mass $m$ of a plate on top of the
  pile. E: When the system collapse particles away from the initial
  cylinder volume (red particles) are weighted to quantify the
  intensity of the collapse. A slice view of the middle of the packing
  is shown to clarify the geometry. See text for details.}
\label{figExpe}
\end{figure*}

We carried out three kinds of experiments, all of which have the same
preparation protocol. The protocol consisted of filling thin-walled
tubes with particles by dropping them one-by-one at a steady rate from
an overhead hopper that was located $40$cm above a base plate, as
shown in Fig.\ref{figExpe}-A). Each tube was a $30$cm tall PVC pipe
whose inner diameter, $\delta$, varied from $2.6$cm to $20.2$cm. The
tubes rested on base plates made of glass for low friction (friction
coefficient between glass and acrylic particle is $\sim 0.5$) or glass
covered with foam for high friction (friction coefficient is higher
than the limit of our experimental measurement method, e.g. $\sim
10$). During the filling process, the system could be vibrated by an
eccentric, driven by a DC motor that was attached to the outside of
the tube at a height of $\sim 1.5 \cdot \delta$. We tuned the speed of
the motor to create a $\sim 1700$m/s$^2$ cyclic acceleration of the
$16$g eccentric. We measured the height, $h$, of the pile inside the
cylinders by gently dropping a disc of paper on top of the particle
packing and measuring the distance between the center of this disc and
the base. We removed this disc before the next step.

As shown in fig.\ref{figExpe}-B, the tube was then carefully and slowly removed by lifting it vertically. Friction between the particles and the cylinder was low enough not to significantly perturb particles inside the cylinder. This was particularly true for the smaller stars, even when the cylinder diameters were small. For the large, 25mm nylon stars, particles could be dragged by the walls, but this was still a relatively rare occurrence, and columns made of them were highly stable during lifting of the tube.

In the first kind of experiment, where the aim was to study the basic
stability of a pile, the next step consisted in measuring the amount
of particles that fell from the column, if any, when the confining
cylinder was removed. Fig.\ref{figExpe}-E shows the method we used to
quantify the degree of the collapse of a pile: after the cylinder was
removed, all particles (red markers in fig.\ref{figExpe}-E) lying
outside the circle made by the intersection of the cylinder and the
base (red circle) were weighed, and their mass was compared to the
total mass of the pile (blue and red markers) to compute a collapse
ratio:

\begin{equation} \label{eqCollapseRatio}
	r=\dfrac{\text{mass of collapsed particles}}{\text{total mass}}
\end{equation}

\noindent
When the pile was stable, $r=0$, and when it was unstable, it approached $1$.

In the second kind of experiment, we studied the stability of the
piles to tilting. A stable system (meaning no, or very few, particles
falling off after the cylinder was removed) was prepared without
vibration, on a foam-covered base (high friction). The base was then
slowly tilted from horizontal to an angle $\theta$ as in
fig.\ref{figExpe}-C. We found that collapses occurred via small
nucleation events, where one or a few particles fell off, which led
quickly to a massive collapse that destroyed the main structure of the
pile. We measured the angles, $\theta$, for the first nucleation event
and the last massive collapse, and computed the collapse ratio, $r$,
for the final event.

A third kind of experiment was carried out to quantify the stability
when the piles were loaded vertically. Like the other experiments, we
first prepared a stable pile on a glass plate without vibration. We
then added a weight to the top (see fig.\ref{figExpe}-D) until it
collapsed. To do so, we first placed a lightweight rigid disc on top
of the pile to distribute the mass. We then placed an empty container
on the disc. Finally, we slowly filled the container with water until
the pile collapsed. We determined the weight, $m$, of the filled
container plus rigid disc at the time of collapse as a function of the
pile height $h$.

To account for the statistical variability, each set experiments was
repeated at least $10$ times to produce an averaged result. All
experimental data presented here are mean values, with error bars
reflecting the variance over trials. In some cases, these error bars
are too small to see. Table \ref{tabExpe} summarizes the different
experiments.

\begin{table*}[htb!]
	\begin{normalsize}
		\begin{center}
			\begin{tabular}{|c|c|c|c|c|c|c|}
			\hline 
			{} & stars & vibration & basis & $\delta$ (cm) & $h$ (cm) & $\mathcal{N}$ \\
			\hline 
			stability & acrylic & no & glass & $4 \to 15.4$ & $2.5 \to 27.5$ & 1 \\ 
			\hline 
			stability & acrylic & yes & glass & $4 \to 15.4$ & $2.5 \to 27.5$ & 2 \\ 
			\hline 
			stability & acrylic & no & foam & $7.7 ; 15.4$ & $2.5 \to 27.5$ & 3 \\ 
			\hline 
			stability & nylon $2.5$cm & no & glass & $4 \to 10.1$ & $6 \to 30$ & 4 \\ 
			\hline 
			stability & nylon $2.5$cm & yes & glass & $4 \to 10.1$ & $6 \to 30$ & 5 \\ 
			\hline 
			stability & nylon $5$cm & no & glass & $7.7 ; 15.4$ & $7 \to 30$ & 6 \\ 
			\hline 
			stability & nylon $10$cm & no & glass & $7.7 ; 15.4$ & $12 \to 30$ & 7 \\ 
			\hline 
			stability & acrylic + beads & no & glass & $13$ & $12.5 \pm 3\%$ & 8 \\ 
			\hline 
			tilting & acrylic & no & foam & $13$ & $8 \to 19.5$ & 9 \\ 
			\hline 
			tilting & nylon $2.5$cm & no & foam & $15.4$ & $19$ & 10 \\ 
			\hline 
			tilting & nylon $5$cm & no & foam & $15.4$ & $19$ & 11 \\ 
			\hline 
			tilting & nylon $10$cm & no & foam & $15.4$ & $19$ & 12 \\ 
			\hline 
			loading & acrylic & no & glass & $13$ & $13.5 \to 18.6$ & 13 \\ 
			\hline 
			\end{tabular} 
		\end{center}
		\caption{Parameters used for the three kinds of experiment: stability, tilting and loading.}
		\label{tabExpe}
	\end{normalsize}
\end{table*}

\section{Results} \label{SecResult}

\subsection{Intrinsic stability of the pile} \label{SecResStab}

\begin{figure*}[htb!]
	\begin{center}
		\includegraphics[width=0.95\textwidth]{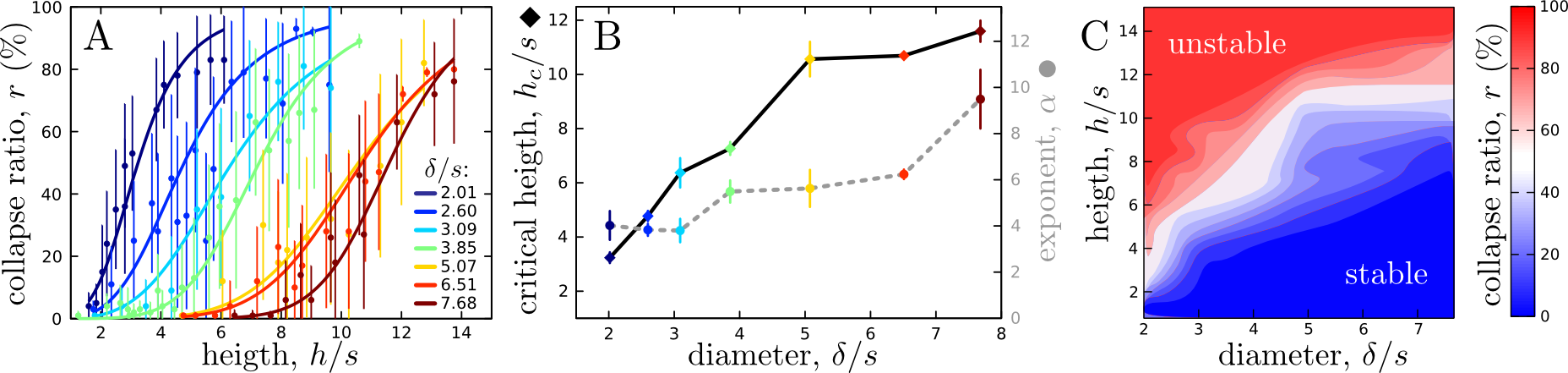}
	\end{center}
\caption{(color online) Stability of the acrylic particles from
  experiment $\mathcal{N}1$. A: ratio $r$ of collapsed particles as a
  function of the pile height $h$ normalized by the particle size,
  $s=2$cm for different heights of acrylic particle piles and
  different cylinder sizes $\delta/s$. For each height, the ratio $r$
  is averaged over at least 10 realizations and an errorbar is
  deduced. Plain curves are the fit of the data with
  eq.\ref{eqFitFormula}. B: Fit parameters $h_c$ (plain black line)
  and $\alpha$ (grey dashed line) for different pile diameters
  $\delta$. C: Stability phase diagram for acrylic particles. The
  ratio $r$ of collapsed particles is plotted as a function of the
  normalized pile height $h/s$ and the normalized pile diameter
  $\delta/s$. Stable domain is where $r$ is the lowest.}
\label{figDiagrmPhas1}
\end{figure*}

In Fig.\ref{figDiagrmPhas1}-A, we present the evolution of $r$ as a
function of $h$ for different cylinder diameters $\delta$. For a given
$\delta$, $r$ varies from $r=0$ (the pile is stable for low height) to
$r \simeq 1$ (the pile is fully collapsed) with a reasonably well
defined transition between the two behaviors. The relative sharpness
of the transition means the range of partially stable systems is very
narrow: in most cases when a part of the pile begins to collapse,
cohesion is lost for the full structure. In order to characterize this
transition and to quantitatively measure the transition height $h_c$
we have fitted the experimental data for $r(h)$ to:

\begin{equation} \label{eqFitFormula}
	r=\dfrac{1} {{\left( \dfrac{h_{c}}{h} \right)}^{\alpha} + 1},
\end{equation}

\noindent where $h_c$ is the height for which $r=1/2$. This expression
indicates a critical height, $h_c$.  For large diameters, $r
\rightarrow 1$ slowly since some particles remain in the original
region covered by the uncollapsed pile (red disc in
fig.\ref{figExpe}-E) after the collapse.

Fitted values of $h_c$ and $\alpha$ for different $\delta$ from experiment $\mathcal{N}1$ are presented in Fig.\ref{figDiagrmPhas1}-B. As suggested qualitatively by Fig.\ref{figDiagrmPhas1}-A, $h_c$ increases almost linearly with $\delta$ and then saturates around a height equivalent to $11$ particle lengths (arm-tip-to-arm-tip along an axis). Fig.\ref{figDiagrmPhas1}-C
shows the extrapolated values of the collapse ratio $r(\delta,h)$ as a
function of $h$, and gives a stability diagram indicating the pile behaviour: stability is greater when $\delta$ is large, and an increase does not necessarily significantly improve the stability.

\begin{figure}[htb!]
	\begin{center}
		\includegraphics[width=0.45\textwidth]{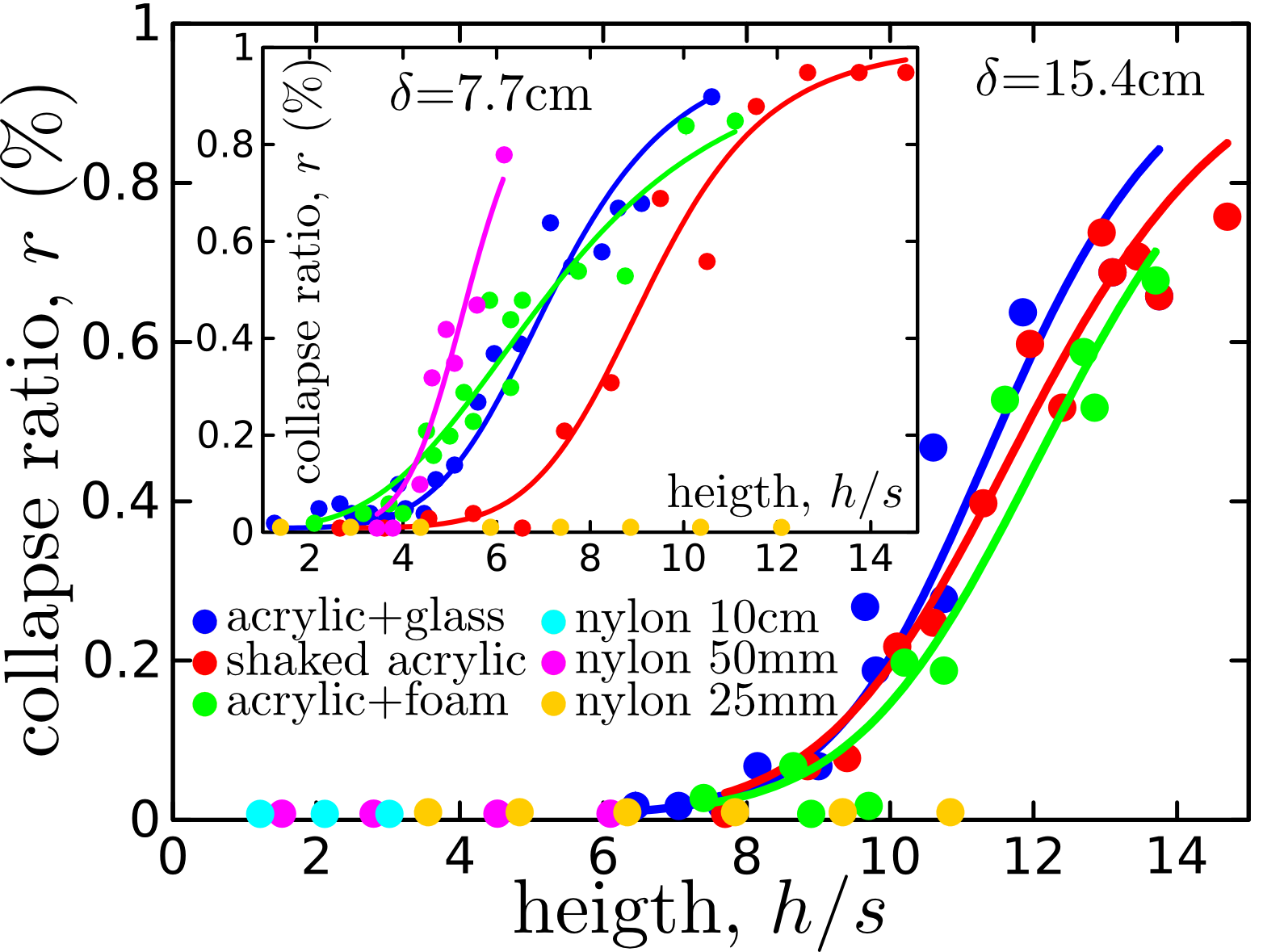}
	\end{center}
\caption{(color online) Ratio $r$ of collapsed particles as a function
  of the scaled cylinder height $h/s$ for two different cylinder
  diameters: $\delta=7.7$cm for the inset graph and $\delta=15.4$cm
  for the main panel. In both cases, experiments were made in
  different situations: (blue) acrylic $2$cm particles on a slippery
  glass base (experiment $\mathcal{N}1$), (red) vibrated acrylic
  particles on glass (experiment $\mathcal{N}2$), (green) acrylic
  particles on a rough foam base (experiment $\mathcal{N}3$),
  (cyan/purple/yellow) nylon $10$/$5$/$2.5$cm particles on glass
  (experiment $\mathcal{N}4,6,7$). Plain curves were fitted with
  eq.\ref{eqFitFormula} and errorbar have been removed by sake of
  clarity.}
\label{figMultiStability}
\end{figure}

We also consider the effect on stability of the particle geometry, the properties of the base, and the preparation. In Fig.\ref{figMultiStability}, we plot $r(h)$ from experiments $\mathcal{N} 1$, $2$, $3$, $4$, $6$ and $7$ for $\delta=7.7$cm and $\delta=15.4$cm, for the following configurations:

\begin{itemize}
	\item $2$cm acrylic particles with the two substrate: slippery glass or rough foam.
	\item $2$cm acrylic particles on glass with or without vibration.
	\item slippery $2$cm acrylic particles or rough $2.5$cm nylon particles. 
	\item nylon particles, changing the arm length: $2.5$cm, $5$cm or $10$cm
\end{itemize}

\noindent Fig.\ref{figMultiStability} shows that there is only a
slight increase of the pile stability when the base is changed from
glass with friction coefficient $\sim 0.5$ to foam with friction
coefficient higher than $10$. Hence, the frictional coefficient of the
substrate does not influence the pile stability
significantly. However, as shown in the inset of
Fig.\ref{figMultiStability}, vibrating the system during the
preparation measurably increases the stability of the pile by shifting
the ratio curve rightward. As shown in the main panel, this effect is
weaker for higher pile diameter. This stabilization effect may be due
to the fact that in the case of larger $\delta$, more particles are
poured, which acts in a similar manner to externally applied
vibration.

Fig.\ref{figMultiStability} shows that increasing the friction
coefficient between particles (from $0.4$ to $1$) dramatically
increases the value of the critical height $h_c$. Note that the size
difference between acrylic ($2$~cm) and nylon ($2.5$~cm) particles is
too small to affect stability. Even with the smaller tube
($\delta=7.7$~cm), we did not observe instability for our tallest
piles ($h_c = 30$~cm) for $2.5$cm nylon particles, whereas $h_c \simeq
14$cm for acrylic particles. We did not observe any stabilisation
effect due to the particle size, but the inset of
Fig.\ref{figMultiStability} shows that $5$~cm nylon particles are less
stable than $2$~cm acrylic stars, even if the friction coefficient is
larger and the arms longer. This is only observed for $\delta=7.7$~cm,
and is due to the fact that the diameter of the pile is comparable to
the particle size. The base of the structure is reduced to too few
particles to be stable.

\begin{figure*}[htb!]
	\begin{center}
		\includegraphics[width=0.9\textwidth]{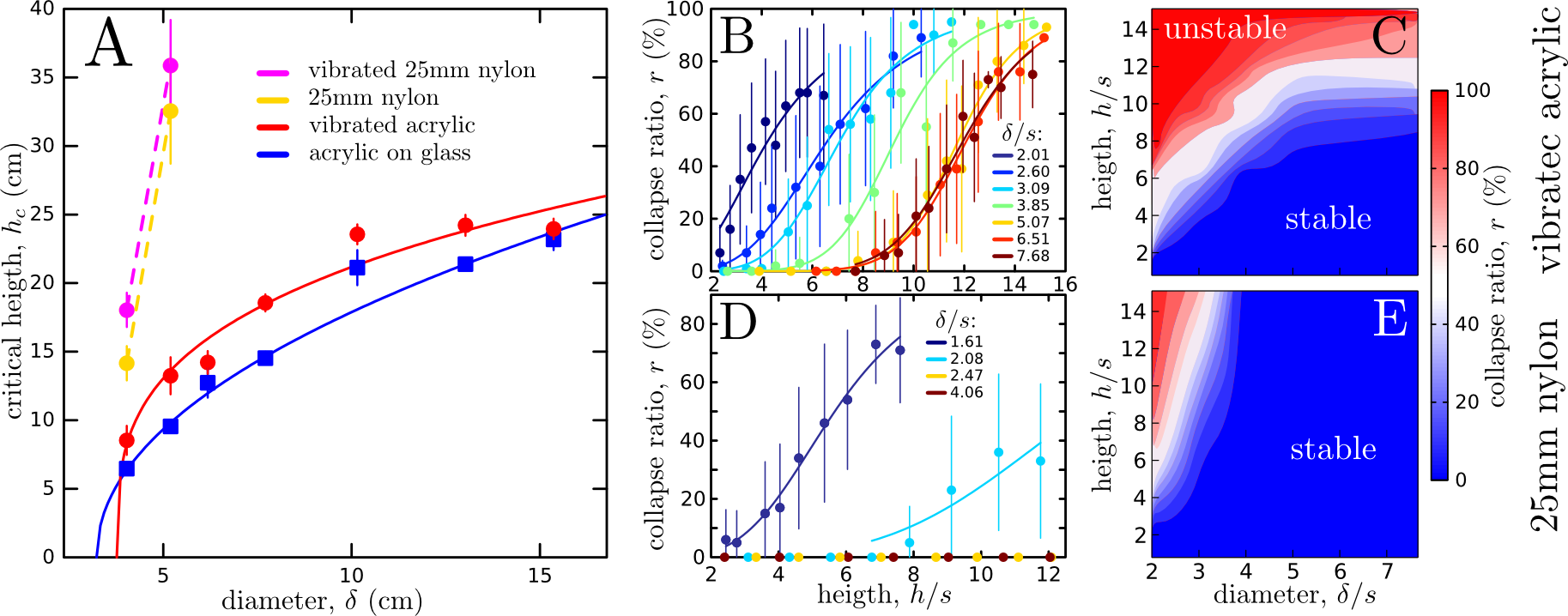}
	\end{center}
\caption{(color online) Effect of vibration and friction on stability,
  for experiments $\mathcal{N}1,2,4,5$ A: Critical stability height
  $hc$ measured from the fitting of eq.\ref{eqFitFormula} as a
  function of the cylinder diameter $\delta$ for different packing
  configurations: vibrated (red) and not vibrated (blue) $2$cm acrylic
  particles on glass, vibrated (pink) and not vibrated (yellow)
  $2.5$cm Nylon particles on glass. Plain lines are curves fitted with
  eq.\ref{eqFitStabity}. B/D: ratio $r$ of collapsed particles as a
  function of the pile height $h$ normalized by the particle size $s$
  for different height of vibrated acrylic/$2.5$cm nylon particle
  piles and different cylinder sizes $\delta/s$. Plain curves are the
  fit of the data with eq.\ref{eqFitFormula}. C/E: Stability phase
  diagram for vibrated acrylic/$2.5$cm nylon partzicles. The ratio $r$
  of collapsed particles is plotted as a function of the normalized
  pile height $h/s$ and the normalized pile diameter $\delta/s$.}
\label{figDiagrmPhase}
\end{figure*}

To more accurately analyse the effect of the particle roughness and
vibration, Fig.\ref{figDiagrmPhase}-C and E show the stability phase
diagram for vibrated piles of $2$cm acrylic particles and $2.5$cm
nylon particles from experiments $\mathcal{N} 2$ and $\mathcal{N} 4$
respectively.  Fits of Eq.\ref{eqFitFormula} to the data yield
$h_c(\delta)$ for these two configurations, as in
Fig.\ref{figDiagrmPhase}-A.

As suggested by Fig.\ref{figMultiStability},
Fig.\ref{figDiagrmPhase}-A shows that vibration improves the pile
stability, even if the effect is weaker for highly frictional particles
and for large $\delta$. Also, this figure confirms that increasing
friction between particles dramatically improves the pile stability,
whatever the pile dimensions, and very significantly shifts the
$h_c(\delta)$ curve upward. To provide a quantitative analysis, we
fitted these curves to:

\begin{equation} \label{eqFitStabity}
	h_c=K(\delta-s_0)^{\beta},
\end{equation}

\noindent where $K$ quantifies the overall stability of the packing,
$\beta$ quantifies the effect of the pile diameter and $s_0$, of the
same order of magnitude as the particle size $s$, gives the smallest
possible diameter. We note that $K$ is larger for vibrated systems and
for rough particles, and that $\beta$ is smaller for vibrated systems
than for non-vibrated ones.

\begin{figure}[htb!]
	\begin{center}
		\includegraphics[width=0.45\textwidth]{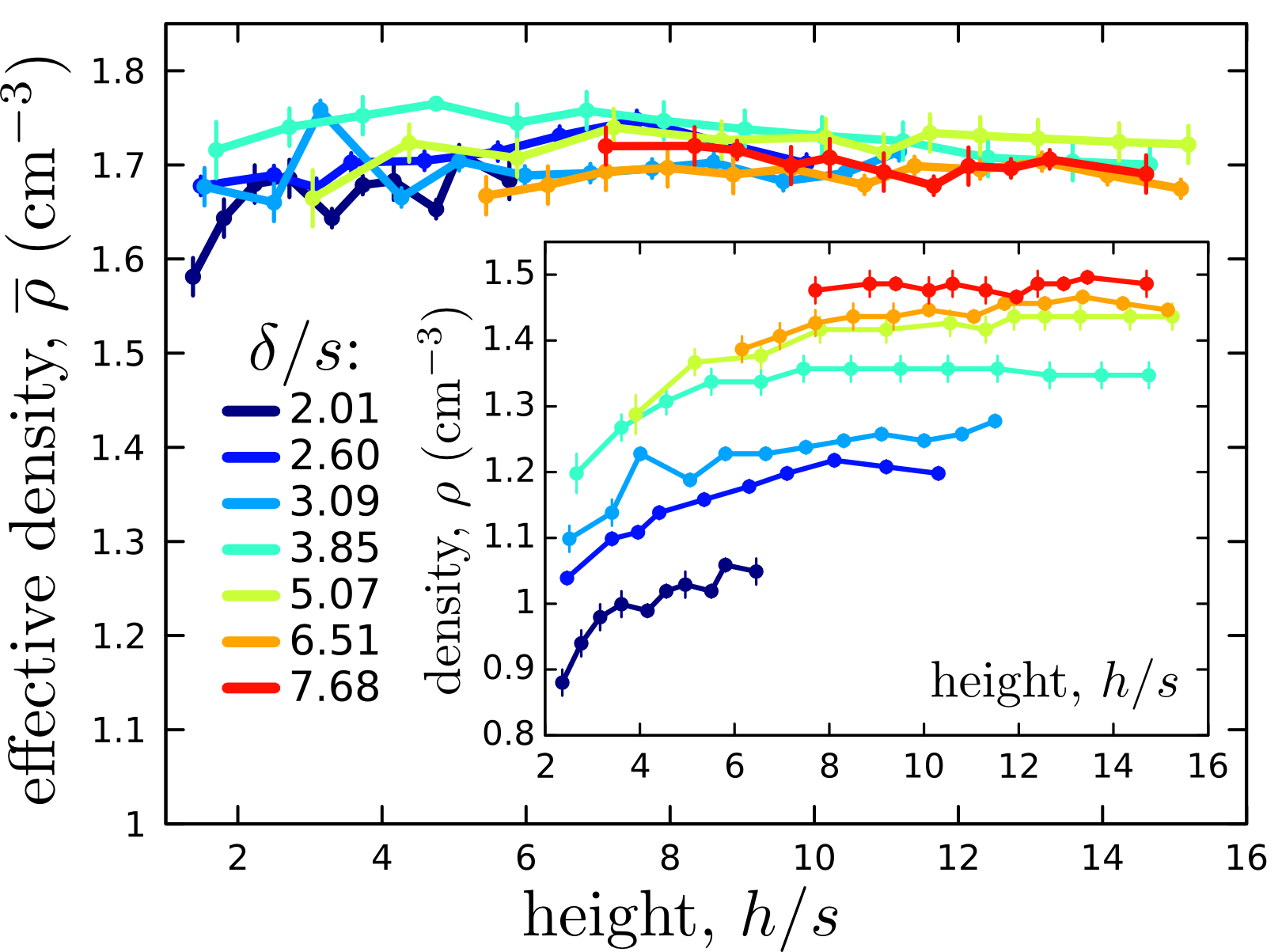}
	\end{center}
\caption{(color online) Inset: Density $\rho$ of vibrated $2$cm
  acrylic particle piles (particles per cm$^3$) as a function of the
  pile height ($h/s$) for different pile diameters. Main panel:
  effective pile density $\overline{\rho}$ measured correcting the
  side effects according to eq.\ref{eqDensity} with
  $\Delta_r=0.75$. Measurements are made from experiment
  $\mathcal{N}2$.}
\label{figDensity}
\end{figure}

To better understand the higher stability of vibrated systems, we
computed the global density, $\rho$, of the $2$cm acrylic particle
piles for the non-vibrated (experiment $\mathcal{N}1$) and vibrated
(experiment $\mathcal{N}2$) systems of different diameters and
different heights. In the inset of Fig.\ref{figDensity} these data are
plotted ($\rho(h,\delta)$) in the case of vibrated piles. The density
seems to vary substantially with $\delta$ and saturate only for large
height. This is due to boundary effects, since the local density is
reduced close to the edges of the cylinder. This can easily be
corrected in eq.\ref{eqDensity} by removing part of volume near the
edges when computing $\rho$:

\begin{equation} \label{eqDensity}
	\overline{\rho}=\dfrac{n}{\pi(h-\Delta_r) \left( \dfrac{\delta-\Delta_r}{2}\right)^2}
\end{equation}
\noindent
In eq.\ref{eqDensity}, $\Delta_r$ quantifies how much of the edge
volume must be removed. It is comparable to an arm length, and is
computed to optimize the collapse of the data in Fig.\ref{figDensity}
($\Delta_r=0.75$cm $\sim 8/3 \cdot $particle size). With this
correction, we find that the density is effectively constant, with an
average value in the non-vibrated case of $\overline{\rho}=1.6 \pm
0.04$cm$^{-3}$, and in the vibrated case of $\overline{\rho}=1.7 \pm
0.03$cm$^{-3}$. Assuming that particle neighbors are located roughly
isotropically, this implies that the average interlocking distance
between particles changed from to $3.70$~mm to $3.57$~mm when the
packing was vibrated. Hence, a variation of the interlocking distance
as small as $0.13$~mm can have a very significant impact on the pile
stability as emphasized by Fig.\ref{figDiagrmPhase}.

\subsection{Destabilization of piles} \label{SecResTilt}

\begin{figure}[htb!]
	\begin{center}
		\includegraphics[width=0.45\textwidth]{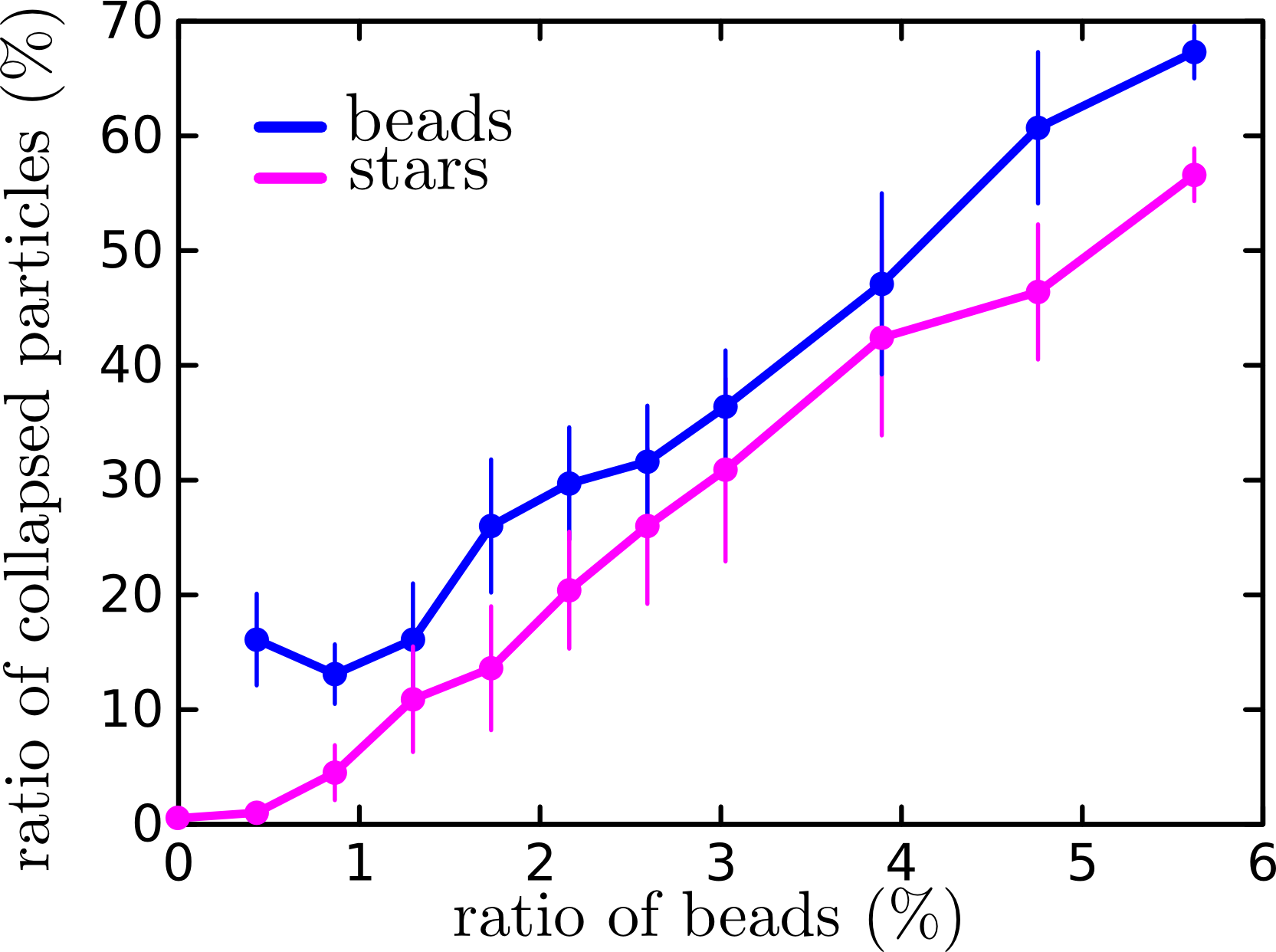}
	\end{center}
\caption{(color online) Fraction of collapsed beads ($0,95$cm diameter
  acrylic beads) to the total number of beads (blue) and ratio of
  collapsed particles ($2$cm acrylic particles) over the total number
  of particles as a function of the ratio of beads in the $2300$
  particles packing. Measurements were made from experiment
  $\mathcal{N}8$.}
\label{figMixbead}
\end{figure}

In order to study other modes of destabilization, we mixed beads among
the particles. The total number of particles in the system was kept
constant ($2300$) but the number ratio between $2$~cm acrylic
particles and $0.95$~cm diameter acrylic beads was varied from $0$ to
$\sim 6\%$ in experiment $\mathcal{N} 8$. We chose $0.95$cm diameter
beads because their radius corresponds to the length of a particle
arm.  Such a spherical particle can be completely interlocked in a
star particle, hence weakening the interlocking of the stars
particles. Also, the bead volume was low enough not to create a strong
height variation as we changed the fraction of beads ($h \approx
12.5$cm). In Fig.\ref{figMixbead}, we show that the addition of
spheres indeed decreases the stability of the pile by linearly
increasing the fraction of collapsed particles to beads in the total
packing. Similarly, the fraction of collapsed beads also increases
linearly with the total fraction of beads. Nevertheless, beads are
always destabilizing, since the collapsed bead fraction is always
higher than the collapsed star ratio.

\begin{figure}[htb!]
	\begin{center}
		\includegraphics[width=0.45\textwidth]{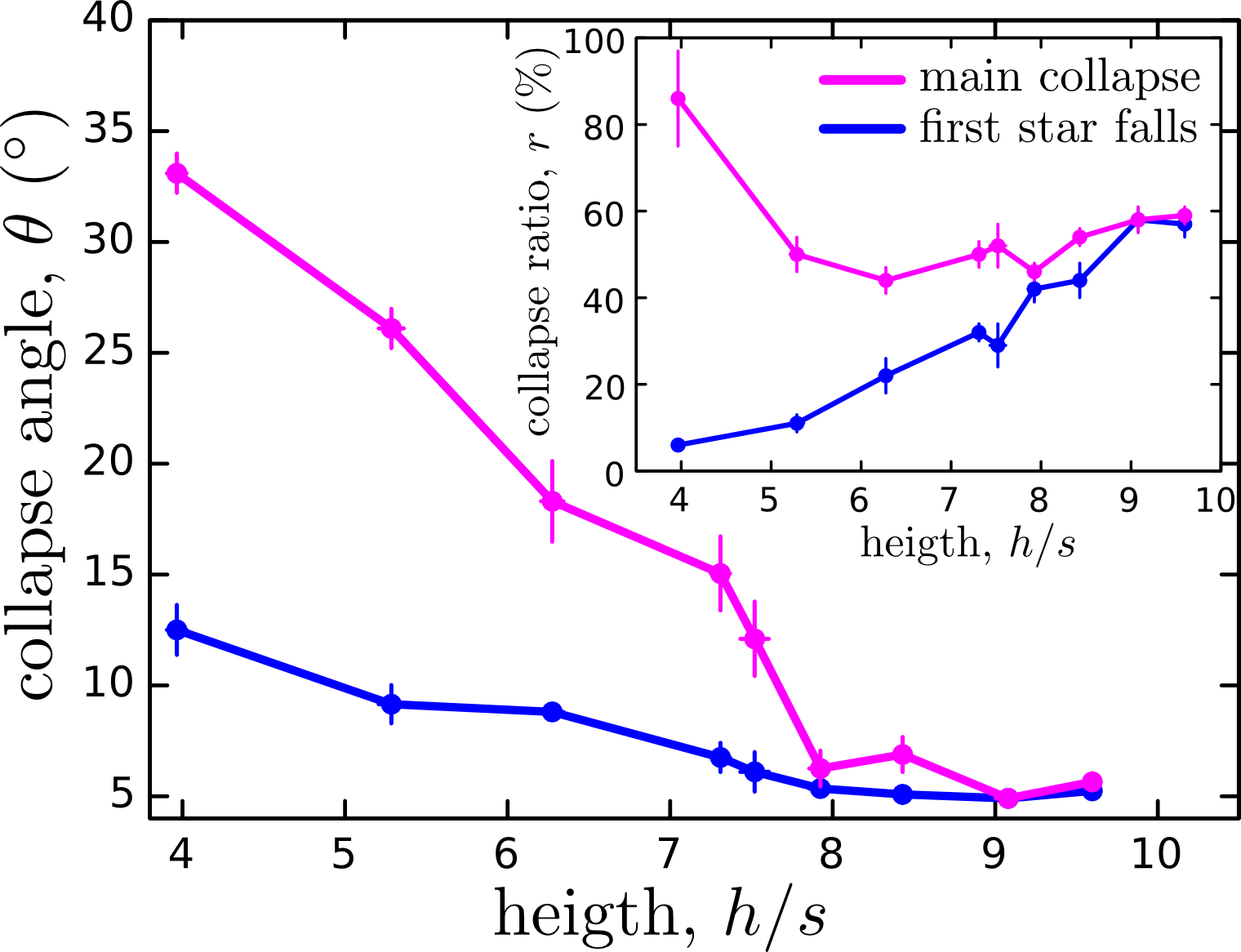}
	\end{center}
\caption{(color online) Tilt angle $\theta$ (see fig.\ref{figExpe}-C)
  for the collapse of the initial particle and for the final collapse
  as a function of the pile height. Inset: Collapse ratio $r$ for
  first and main collapse. Experiments were carried out with $2$cm
  acrylic particles in a $13$cm diameter cylinder ($\mathcal{N}9$). }
\label{figTilting}
\end{figure}

Stable systems can also be unbalanced by tilting. Slowly tilting a
stable pile of $2$cm acrylic particles induced an initial small
collapse of a few particles, corresponding to a small initial
rearrangement of the pile. With increasing tilt, this was followed by
a series of other small collapses, leading to a final major failure
that was system-spanning. The tilting experiment, $\mathcal{N}9$, was
carried out with a $13$cm diameter tube on a rough foam substrate for
various $h$. Fig.\ref{figTilting} presents the evolution of the
initial and final collapse angles, $\theta(h)~vs.~h$. The initial
collapse occurred early in the tilting process and is clearly
separated from the last one for low enough $h$. But, for higher $h$,
the gap between initial and final collapse angles coalesced, and there
was only one large collapse. The inset of Fig.\ref{figTilting} shows
the ratio of particles lost by the pile after the first and last
collapses. As $h$ increased, the angular gap between the two collapses
decreased and the ratios of collapsed particles converged, $i.e.$ the
first and last collapses converge in both angle and intensity. Also,
as expected, whether the criterion of stability is the first or last
collapse angle, the critical angle rapidly decreased with the pile
height.

We also carried out tilting experiments with fixed pile sizes
($\delta=15.4$cm and $h=19$cm) and varying particle sizes. We used
nylon particles with arm lengths $s=2.5$cm, $5$cm and $10$cm
(experiments $\mathcal{N}10,11,12$ respectively). Due to their high
friction, nylon particles tended to fall over as a rigid body at the
collapse angle, corresponding to an imbalance of the pile center of
mass. The final angles of collapse are given in
Table.\ref{tabExpeTilt} which shows that the longer the arms, the less
stable the pile. This is due to the fact that the larger the particle,
the less contact points there were between the pile and the base and
the lower the stability.

\begin{table}[htb!]
	\begin{footnotesize}
		\begin{center}
			\begin{tabular}{|c|c|}
			\hline 
			particle size (cm) & collapse angle ($^\circ$) \\ 
			\hline 
			2.5 & 29.2 \\ 
			\hline 
			5 & 25.0 \\ 
			\hline 
			10 & 14.7 \\ 
			\hline 
			\end{tabular} 
		\end{center}
		\caption{Angle of collapse for piles with $h=19$cm and $\delta=15.4$cm and different nylon particle size. Results from experiments $\mathcal{N}10,11,12$.}
		\label{tabExpeTilt}
	\end{footnotesize}
\end{table}

\begin{figure}[htb!]
	\begin{center}
		\includegraphics[width=0.45\textwidth]{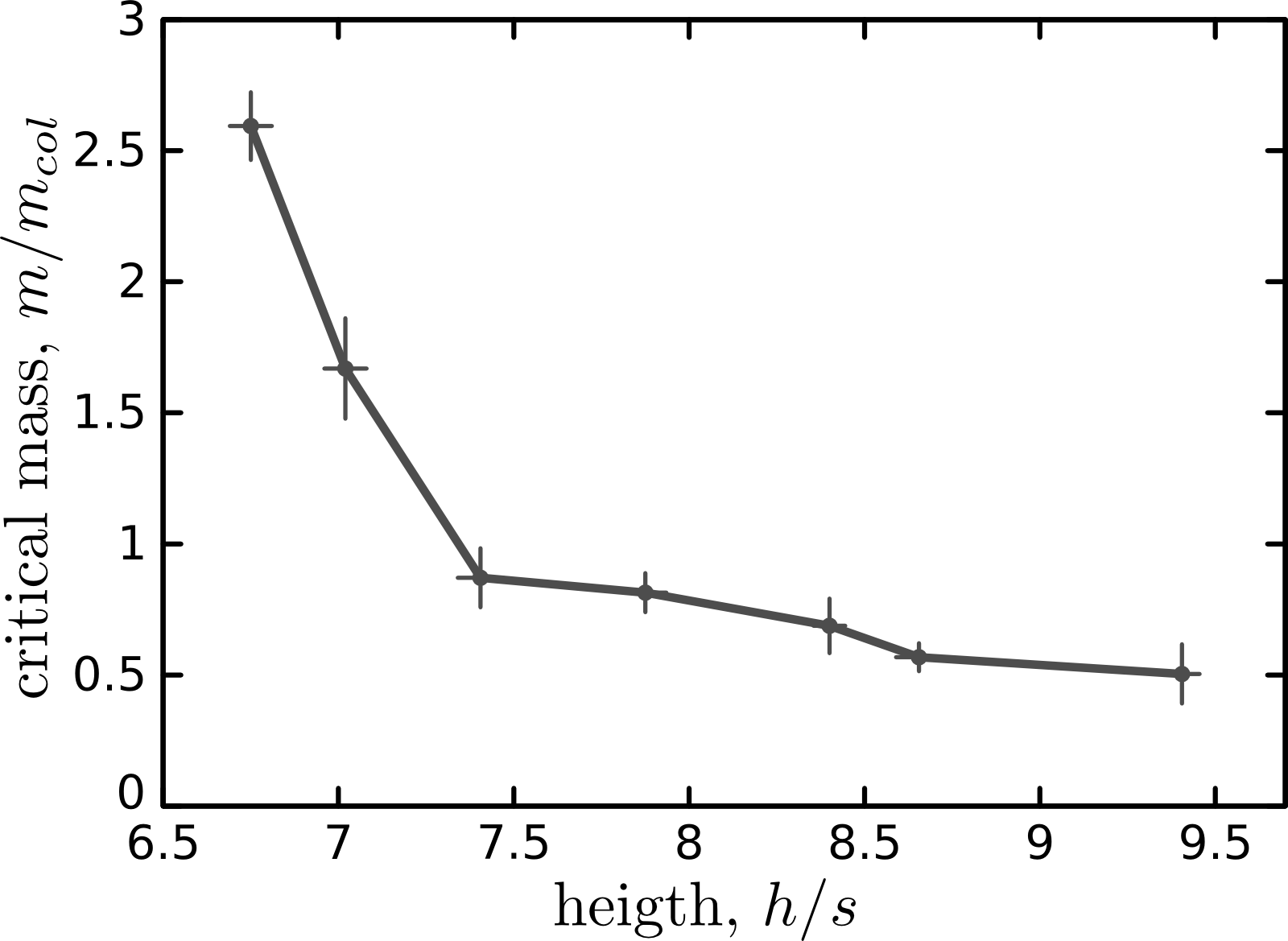}
	\end{center}
\caption{Maximum mass, $m$, supported by a stable $13$cm diameter pile
  before collapsing, as a function of the scaled pile height
  ($h/s$). The loading is given as a fraction of the pile mass. Non
  vibrated $2$cm acrylic particles were used on a glass base
  (experiment $\mathcal{N}13$).}
\label{figMass}
\end{figure}

In the third and last exploration, we prepared piles and then loaded
them vertically. Following our previous preparation protocol, we
prepared stable piles of $2$cm acrylic particles in a tube of $13$cm
diameter, and slowly loaded them vertically. In this case, the
collapse happened as a single event at a critical loading mass
$m$. Data for $m~vs.~h$ in Fig.\ref{figMass} shows that $m$ decreases
sharply with $h$. Also, we notice comparing this graphe with data from
Fig.\ref{figDiagrmPhase}-A that the mass to make a pile collapse is
higher than the mass of particle you should add to make it collapsed
because it is too high. For example, considering the case where
$m=m_{col}$ (for $h/h_s \approx 7.4$), if the stabilization process
was just a matter of vertical loading, this would mean that $h_c$
should be $hc/s \approx 15$, whereas Fig.\ref{figDiagrmPhas1}
indicates that it is $hc/s \approx 11$. This implies that a pile does
not collapse because it cannot support its own weight, but because of
low cohesion of the particle when the pile becomes too high.

\section{Discussion and Conclusions} \label{SecDiscussions}

We have shown experimentally that star-shaped particles can form
stable cylindrical aggregates whose height can be more than three time
larger than the base diameter. The stability of these aggregates was
tested with and without loading and tilting for particles covering a
broad range of arm sizes and frictional properties. For a given type
of star, the pile critical height, $h_c$ rapidly increased with the
diameter of the base, $\delta$, when the latter was comparable to the
particle size, but plateaus for larger diameters. This means that to
increase the height of a cylindrical aggregate one can adjust the base
diameter somewhat, but it is more effective to increase the particle
size to keep it just smaller than $\delta$. There are other ways to
increase the critical height. First, by increasing the friction of the
base, the first particle layer is stabilized, which in turn,
stabilizes the whole structure. In general, fixing the first layer is
a good way to increase the stability of the whole aggregate. Second,
by vibrating the system while raining in the particles, the packing
fraction increases, and the distance between particles decreases which
also enhances stability. This is particularly the case if the system
is small, because less shaking energy is required to create
rearrangements. Thus, not only is the intrinsic geometry of the system
important, but also, the preparation method can improve the aggrate
stability. Third, as long as the particle size remains significantly
smaller than the pile diameter, using larger particles also increases
the stability. But, this effect reverses when the particles become
comparable to the cylinder diameter because the effective base of the
pile (diameter minus arm length) vanishes. Fourth and finally, the
most important parameter for stabilizing a packing, keeping other
properties constant, is the inter-particle friction coefficient. For
instance, we find for a given cylinder diameter and 
particle friction coefficient $\sim 1$, that a stable aggregate can be
more than two times higher than for a friction coefficient of
$0.4$. Hence, the choice of material or surface properties for the
particles for the present experiments is certainly the most important
parameter to obtain a tall stable aggregate.

Star-shaped particle aggregates that are stable initially, can sustain
destabilization effects, within limits. First, adding a certain
fraction of beads with the stars leads to a less stable packing, then
when there are no beads. Conversely, it follows that adding convex
particles to spheres can help stabilize the packing, although we have
not investigated the high sphere fraction limit. Second, stable
cylindrical packings collapse via a succession of avalanches when
slowly tilted. The collapse angles vary nearly linearly with pile
height. This suggests the possibility of creating structures with more
complex aggregate geometries, than purely cylindrical and
vertical. Third, and finally, a stable packing of star-shaped
particles can support up to twice its own weight. However, the ability
to support added weight decreases rapidly with pile height.

The results reported in this work represent a step forward in the
knowledge of both architectural design and granular science. It gives
a better understanding of the self-sustained stability of non-convex
3D particles and their ability to form vertical structures, while it
permits a widening of possible design applications. In addition to the
scientific aspects, we believe this work constitutes a ``handbook'' of
mechanical rules to improve the design of aggregated structures by
giving methods to make them stronger, more stable and more reliable.


\bibliographystyle{unsrt}
\bibliography{ArticleV6.bib}

\end{document}